\documentclass[aps,prl,twocolumn,groupedaddress]{revtex4}

\usepackage[dvipdfm]{graphicx}
\begin{document}
\bibliographystyle{prsty}
\title{Depth profile photoemission study of thermally diffused Mn/GaAs (001) interfaces}

\author{Y.~Osafune}
\affiliation{Department of Complexity Science and Engineering and Department of Physics, University of Tokyo, 
7-3-1 Hongo, Bunkyo-ku, Tokyo, 113-0033, Japan}
\author{G.~S.~Song}
\affiliation{Department of Complexity Science and Engineering and Department of Physics, University of Tokyo, 
7-3-1 Hongo, Bunkyo-ku, Tokyo, 113-0033, Japan}
\author{J.~I.~Hwang}
\affiliation{Department of Complexity Science and Engineering and Department of Physics, University of Tokyo, 
7-3-1 Hongo, Bunkyo-ku, Tokyo, 113-0033, Japan}
\author{Y.~Ishida\footnote{Present adress: RIKEN SPring-8 Center, Sayo-gun, Hyogo 679-5148, Japan}}
\affiliation{Department of Complexity Science and Engineering and Department of Physics, University of Tokyo, 
7-3-1 Hongo, Bunkyo-ku, Tokyo, 113-0033, Japan}
\author{M.~Kobayashi}
\affiliation{Department of Complexity Science and Engineering and Department of Physics, University of Tokyo, 
7-3-1 Hongo, Bunkyo-ku, Tokyo, 113-0033, Japan}
\author{K.~Ebata}
\affiliation{Department of Complexity Science and Engineering and Department of Physics, University of Tokyo, 
7-3-1 Hongo, Bunkyo-ku, Tokyo, 113-0033, Japan}
\author{Y.~Ooki}
\affiliation{Department of Complexity Science and Engineering and Department of Physics, University of Tokyo, 
7-3-1 Hongo, Bunkyo-ku, Tokyo, 113-0033, Japan}
\author{A.~Fujimori\footnote{E-mail: fujimori@wyvern.phys.s.u-tokyo.ac.jp}}
\affiliation{Department of Complexity Science and Engineering and Department of Physics, University of Tokyo, 
7-3-1 Hongo, Bunkyo-ku, Tokyo, 113-0033, Japan}
\author{J.~Okabayashi\footnote{Present adress: Department of Physics, Tokyo Institute of Technology, 2-12-1 Ookayama, Meguroku, Tokyo 152-8551, Japan}}
\affiliation{Department of Applied Chemistry, University of Tokyo,
7-3-1 Hongo, Bunkyo-ku, Tokyo, 113-8656, Japan}
\author{K.~Kanai}
\affiliation{Department of Applied Chemistry, University of Tokyo,
7-3-1 Hongo, Bunkyo-ku, Tokyo, 113-8656, Japan}
\author{K.~Kubo}
\affiliation{Department of Applied Chemistry, University of Tokyo,
7-3-1 Hongo, Bunkyo-ku, Tokyo, 113-8656, Japan}
\author{M.~Oshima}
\affiliation{Department of Applied Chemistry, University of Tokyo,
7-3-1 Hongo, Bunkyo-ku, Tokyo, 113-8656, Japan}
\date{\today}
\
\begin{abstract}
We have performed a depth profile study of thermally diffused Mn/GaAs (001) interfaces using photoemission spectroscopy combined with Ar$^+$-ion sputtering. We found that Mn ion was thermally diffused into the deep region of the GaAs substrate and completely reacted with GaAs. In the deep region, the Mn 2$p$ core-level and Mn 3$d$ valence-band spectra of the Mn/GaAs (001) sample heated to 600 $^{\circ}$C were similar to those of Ga$_{1-x}$Mn$_x$As, zinc-blende-type MnAs dots, and/or interstitial Mn in tetrahedrally coordinated by As atoms, suggesting that the Mn atoms do not form any metallic compounds but are tetrahedrally coordinated by ligand atoms and Mn 3$d$ states are hybridized with ligand orbitals. Ferromagnetism was observed in the dilute Mn phase.
\end{abstract}

\maketitle

\section*{I. INTRODUCTION}
Since the discovery of ferromagnetism in Mn-doped III-V semiconductors such as In$_{1-x}$Mn$_x$As \cite{Munekata} and Ga$_{1-x}$Mn$_x$As \cite{Ohno_96}, diluted magnetic semiconductors (DMS's) have been extensively studied because these materials are promising candidates for applications in ``spintronics" (spin electronics) semiconductor devices. Ferromagnetism in the III-V-based DMS's is considered to be caused by coupling between the local magnetic moments of the magnetic ions mediated by charge carriers of the host semiconductors, that is, ``carrier-induced ferromagnetism". The prototypical III-V DMS Ga$_{1-x}$Mn$_x$As has been most extensively studied by many reserchers after the discovery of the relatively high Curie temperatures ($T\mathrm{_C}$ $\sim$110 K) \cite{Matsukura} and characteristic magnetotransport properties such as giant magnetoresistance and anomalous Hall effect \cite{Oiwa, Esch}. The Mn doping into GaAs has been achieved by low temperature molecular beam epitaxy (MBE) method. Recently, ferromagnetism has been observed in Mn/GaN and Cr/GaN interfaces fabricated by thermal annealing of Mn and Cr films deposited on GaN substrate \cite{GaN}. 
%A fabrication of DMS by thermal diffusion method, which has played an important role in semiconductor device fabrication, 
In such a system, because the density and the chemical state of the thermally diffused ions vary as functions of the depth, it is important to investigate the depth profile of the distribution, chemical composition, and chemical bonding of the diffused ion.

Photoemission spectroscopy (PES) is a powerful technique to elucidate the electronic structure of solids, for instance, detailed electronic structure of Ga$_{1-x}$Mn$_x$As has been clarified by PES \cite{Okabayashi_PES, Okabayashi_RPES, Okabayashi_ARPES, Okabayashi_ARPES2}. Furthermore, PES combined with Ar$^+$-ion sputtering is suitable for depth profile analysis in studying diffusion effect \cite{Ishida, Hwang}. For the above reasons, in this work, we have conducted Mn doping by thermal annealing of Mn film deposited on a GaAs substrate and performed the depth profile study of thermally diffused Mn/GaAs (001) interfaces using PES combined with Ar$^+$-ion sputtering to investigate the electronic structure of Mn/GaAs (001) interfaces from shallow to deep regions. 

\begin{figure}[b]
%\begin{center}
\includegraphics[width=8cm]{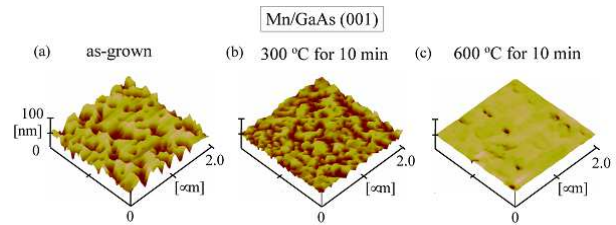}
\caption{(Color online) Atomic force microscopy images for (a) as-grown sample, (b) sample post-annealed up to 300 $^{\circ}$C for 10 min, and (c) sample post-annealed up to 600 $^{\circ}$C for 10 min.}
\label{AFM}
%\end{center}
\end{figure}

\section*{II. EXPERIMENT}
Thermally diffused Mn/GaAs (001) interfaces were fabricated by the MBE method as follows. First, an ordered GaAs buffer layer with 170 nm thickness was grown on an epi-ready Si-doped $n^{+}$-GaAs (001) substrate with carrier concentration $\sim$10$^{20}$ cm$^{-3}$. During the Ga and As deposition, the GaAs substrate was heated up to 600 $^{\circ}$C . Then, Mn metal was deposited on the GaAs buffer layer to $\sim$2 nm thickness at the rate of 0.029 nm/s at the substrate temperature of 50 $^{\circ}$C. After the Mn deposition, Mn diffusion into the GaAs substrate was achieved by post-annealing. During the annealing, As flux of $2.5\times10^{-5}$ Torr was supplied when the substrate temperature exceeded 400 $^{\circ}$C to prevent As deficiency. Figure 1(a)-(c) shows the surface morphology of the as-grown and annealed (300 $^{\circ}$C and 600 $^{\circ}$C) Mn/GaAs (001) samples measured by atomic force microscopy (AFM) in air. The AFM images suggested that the level of surface roughness decreased with increasing temperature. In particular, a fairly flat surface was obtained for the thermally diffused Mn/GaAs (001) annealed at 600 $^{\circ}$C. Surface reconstruction was monitored during the growth and the post-annealing with reflection high-energy electron diffraction (RHEED) and the RHEED image showed a c(4$\times$4) pattern, which is one of the reconstructed surface of GaAs (001) in the As-rich condition as reported in Ref \cite{surface}, after the post-annealing up to 600 $^{\circ}$C for 10 min. The depth profile studies were performed on the 600 $^{\circ}$C-annealed sample. For the PES measurements, the 600 $^{\circ}$C-annealed sample was capped by amorphous As with the thick of less than 5 nm to prevent oxidation of the sample surface during transfer of the sample in air from the MBE chember to the PES spectrometer. Prior to the PES measurements, the As cap was removed by sputtering.

Ultraviolet photoemission spectroscopy (UPS) measurements were performed at beamline BL-18A of Photon Factory (PF), Institute for Material Structure Science, High Energy Accelerator Research Organization (KEK) using a VG CLAM hemispherical analyzer. All the photoemission spectra were taken at room temperature under an ultra high vacuum of 9.0 $\times$ 10$^{-10}$ Torr. The Fermi level ($E\mathrm{_F}$) was calibrated using the Fermi edge of a Cu metal in electrical contact with the sample. X-ray photoemission spectroscopy (XPS) measurements were performed at the University of Tokyo using a VSW125 analyzer. The total energy resolution including temperature broadening was estimated to be $\sim$200 meV and $\sim$800 meV for UPS and XPS, respectively. In both UPS and XPS measurements, photoelectrons were collected in the angle-integrated mode. Depth profile studies were performed by repeated cycles of sputter-etching and subsequent PES measurement. Sputter-etching was done with Ar$^+$-ion at 1.0 kV. The sputtering rate was approximately 0.1-0.2 nm/min for XPS and 0.2-0.4 nm/min for UPS. 
The relative chemical compositions of Ga, As, and Mn on the surface and in deep region were estimated from the peak intensity using the equation $C_{x}$ = ($I_{x}$/$S_{x}$)/($\Sigma$$I_{i}$/$S_{i}$), where the $I_{x}$ is the peak intensity of a core-level $x$ and $S_{x}$ = $\lambda_{x}\cdot\sigma_{x}\cdot$$F_{x}$ ($\lambda_{x}$ : mean free path in GaAs \cite{Mfp}, $\sigma_{x}$ : photoionization cross section of $x$ \cite{Lindau}, $F_{x}$ : instrumental sensitivity factor of $x$).
Magnetization measurements were performed using a SQUID magnetometer (MPMS, Quantum Design, Co., Ltd.).

\begin{figure}[tbp]
\begin{center}
\includegraphics[width=7.6cm]{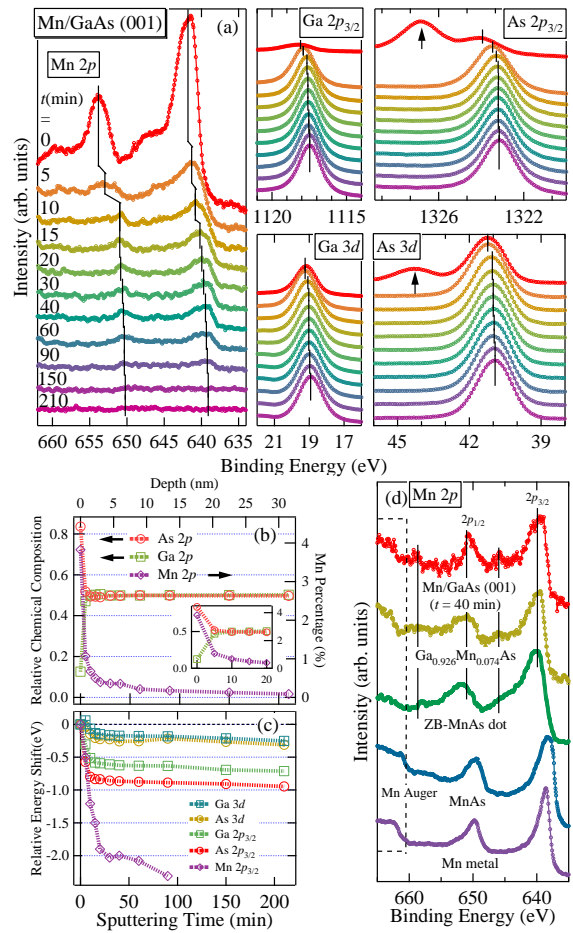}
\caption{(Color online) Core-level XPS spectra in the sputter-etching series of the Mn/GaAs (001) sample post-annealed up to 600 $^{\circ}$C for 10 min. (a) Mn 2$p$, Ga 2$p_{3/2}$, As 2$p_{3/2}$, Ga 3$d$, and As 3$d$ core-level spectra. $t$ (min) denotes the sputtering time. Arrows indicate As core-level signals from oxidized capping layer. Vertical lines represent a peak position of each core level. (b) Relative chemical compositions of Ga, As, and Mn estimated from the peak area of the Ga 2$p_{3/2}$, As 2$p_{3/2}$, and Mn 2$p_{3/2}$ core levels as functions of sputtering time (bottom) and depth (top). The sputtering rate was 0.1-0.2 nm/min. Depth was estimated by multiplying the sputtering time by the average sputtering rate of 0.15 nm/min. The inset shows an enlarged plot upto $t$ = 20 min. (c) Relative peak shifts of the Ga 3$d$, As 3$d$, Ga 2$p_{3/2}$, As 2$p_{3/2}$, and Mn 2$p_{3/2}$ core levels as functions of sputtering time. (d) Comparison of the Mn 2$p$ core-level spectra between the thermally diffused Mn/GaAs (001) interface sputtered for $t$ = 40 min, Ga$_{0.926}$Mn$_{0.074}$As \cite{Okabayashi_PES}, ZB-type MnAs dots \cite{Ono_MnAs}, hexagonal MnAs, and Mn metal. Mn $L_{3}M_{2,3}V$ Auger signals are also indicated by dashed lines.}
\label{XPS}
\end{center}
\end{figure}

\section*{III. RESULTS AND DISCUSSION}
%\subsection{III. X-RAY PHOTOEMISSION SPECTROSCOPY}
Figure 2(a) and (b) shows the core-level XPS spectra of an Ar$^+$-ion sputter-etching series for the sample post-annealed up to 600 $^{\circ}$C for 10 min and  their relative core-level intensities as functions of sputtering time ($t$), respectively. The Mn 2$p$ peak completely disappeared at $t$ = 210 min, indicating that the GaAs substrate appeared at $t$ = 210 min. 
%We therefore consider that the intensity ratio of Ga 2$p_{3/2}$ to As 2$p_{3/2}$ at this stage corresponds to the atomic ratio Ga : As = 1 : 1. 
Throughout the sputter-etching series, the Mn 2$p$ spectra did not show line shapes of Mn metal nor metallic Mn compounds. The observation indicates that the Mn metal layer entirely reacted with the GaAs substrate and that Mn was not metallic even in the first surface layer. In the initial stage of the sputtering ($t$ = 0-10 min), we observed the extra As core-level peaks with excess intensity at a higher binding energy ($E_B$) [Fig.\,2(a)] and the $E_B$ of the Mn 2$p_{3/2}$ peak position was relatively high compared with that of after $t$ = 10 min [Fig.\,2(c)]. 
These may be attributed to the influence of As-capping layer and the oxidation of the capping layer and Mn underneath during the transfer of the sample from the MBE chamber to the spectrometer. One can see that the As intensity was rapidly decreased and the Mn 2$p_{3/2}$ position was rapidly shifted to lower $E_B$ for $t$ = 0-20 min by sputtering. 
The Mn 2$p$ spectrum at $t$ = 0 min has prominent structures on the higher $E_B$ side of the 2$p_{3/2}$ and 2$p_{1/2}$ peaks, which may be attributed to the satellite structure of surface Mn oxides. 
Since the Ga-to-As intensity ratio became constant after 20 min sputtering, it is probable that the As-cap layer was removed and the Mn diffusion layer appeared from this point. 
Except for the surface region, the intensities of the Mn 2$p_{3/2}$ peak changed so slowly, indicating that Mn atoms were diffused deep into the substrate where a dilute Mn phase was formed. 
Additionally,  the presence of Ga 2$p$ at $t$ = 0 min may be due to Ga atoms diffused into the As cap layer considering the small photoelectron escape depth for Ga 2$p$ XPS (If we assume the thickness of the As cap layer to be 3 nm and only Mn atoms diffused into the As cap layer, the relative intensity of Ga 2$p$ would be as small as $\sim$0.032 which is much smaller than the observed Ga 2$p$ intensity of $\sim$0.13.). %The Ga diffusion up into the As cap layer conjures the image of the formation of Ga vacancies at the reacted region.
%This observation is consistent with the large diffusion coefficient for the thermal diffusion of Mn at Mn/GaAs interfaces studied by Rutherford backscattering \cite{Hilton_04, Hilton_05}. 

Figure 2(d) shows the Mn 2$p$ spectrum of Mn/GaAs (001) sputtered for $t$ = 40 min compared to those of Ga$_{0.926}$Mn$_{0.074}$As \cite{Okabayashi_PES}, zinc-blende-type (ZB-type) MnAs dots \cite{Ono_MnAs}, hexagonal MnAs, and Mn metal. As for Mn/GaAs (001), each of the spin-orbit component had a main peak and a strong shoulder structure, that is, a charge-transfer satellite located on the higher $E_B$ side of the main peak. The existence of the charge-transfer satellite indicates that there are strong Coulomb interaction between the Mn 3$d$ electrons and strong hybridization between the Mn 3$d$ and other valence orbitals. This structure is similar to that of Ga$_{0.926}$Mn$_{0.074}$As or ZB-type MnAs dots as shown in Fig.\,2(d). Therefore, we concluded that the Mn atoms in the deep region of Mn/GaAs(001) are situated in a chemical environment slimilar to that of Mn in Ga$_{1-x}$Mn$_x$As and ZB-type MnAs dots.

\begin{figure}[tbp]
\begin{center}
\includegraphics[width=8cm]{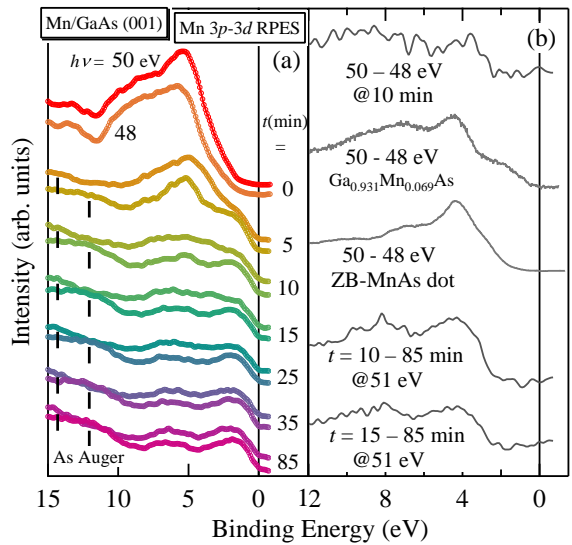}
\caption{(Color online) Mn 3$p$-3$d$ RPES spectra of the Mn/GaAs(001) sample post-annealed up to 600 $^{\circ}$C. (a)Valence-band spectra of the sputter-etching series taken at $h\nu =$ 50 eV (one-resonance) and 48 eV (off-resonance). Vertical bars represent the As Auger signal. The sputtering rate was 0.2-0.4 nm/min. (b) Comparison of the Mn 3$d$ PDOS of the thermally diffused Mn/GaAs (001) sample with that of Ga$_{0.931}$Mn$_{0.069}$As \cite{Okabayashi_RPES} and ZB-type MnAs dots \cite{Okabayashi_MnAs}. The Mn 3$d$ PDOS were also deduced from the difference between the spectra at $t$ = 10 or 15 min and 85 min taken at $h\nu$ = 51 eV.}
\label{PDOS}
\end{center}
\end{figure}

%\subsection{RESONANT PHOTOEMISSION SPECTROSCOPY}
Figure 3(a) shows a series of valence-band spectra for various sputtering times taken with photon energies in the Mn 3$p$-3$d$ core-excitation region. The intensities have been normalized to the Au mirror current. Throughout the depth profile study, Mn-derived Auger signal was not observed and the spectral intensity is suppressed near $E\mathrm{_F}$. These results are characteristics of localized Mn 3$d$ states and are different for metallic Mn compounds \cite{Ishida}, for which strong Auger signal and Fermi edge are observed. Although the valence-band spectra dramatically changed during the first 10 minutes, there was little change after that. Hence, we conclude that the spectra of the first 10 minutes correspond to those of As-cap layer and the subsequent spectra represent those of the dilute Mn phase. The resonantly enhanced Mn 3$d$ partial density of states (PDOS) has been obtained by subtracting the off-resonant ($h\nu =$ 48 eV) spectrum from the on-resonant ($h\nu =$ 50 eV) one as shown at the top of panel\,(b). In order to properly carry out the subtraction, we took into account the photon energy dependence of the photoionization cross-section of As 4$p$. There was a main peak at $E_B$ $\sim$3.7 eV and a broad charge transfer satellite at $E_B$ = 5-13 eV in the Mn 3$d$ PDOS at $t$ = 10 min, similar to Ga$_{1-x}$Mn$_x$As \cite{Okabayashi_RPES} as well as to ZB-type MnAs dots (spectra of the ``medium density" in Ref \cite{Okabayashi_MnAs}). The difference between on-resonance and off-resonance spectra completely disappeared at $t$ = 85 min, and thus we consider that the Mn diffusion layer becomes below the detection limit at $t$ = 85 min. The absence of Mn at $t$ = 85 min was also confirmed by the absence of Mn 2$p$ core-level signal for this series of sputtering (not shown). Here, we should note that the sputtering time $t$ does not indicate the same depths for the XPS and UPS spectra, because the sputtering rate during the UPS measurement was about 2 times higher than that of XPS. We have also attempted to obtain the Mn 3$d$ PDOS by subtracting the spectrum at $t$ = 85 min (GaAs) from the one at $t$ = 10 min, 15 min (Mn/GaAs (001) interface in the dilute Mn phase) for the fixed photon energy of $h\nu$ = 51 eV as shown at the bottom of Fig.\,3(b). These difference spectra are similar to those obtained from the RPES measurement. 

It is supposed that these obtained Mn 2$p$ XPS and 3$d$ PDOS contain some contribution from interstitial Mn atoms (Mn$_\textup{\scriptsize{I}}$). Recently, Jakie\l a  $et$ $al$. proposed that the Mn atoms are diffused into GaAs through interstitial sites \cite{Mndiffusion}. In the zinc-blende lattice there are three possible interstitial sites, two tetrahedral sites, which are surrounded by four cations or four anions, and one hexagonal site \cite{Blinowski}. Because Mn$_\textup{\scriptsize{I}}$ acts as a double donor and is positively charged (Mn$_\textup{\scriptsize{I}}^{2+}$), Mn$_\textup{\scriptsize{I}}$ is electrically stable near the anions \cite{Edmonds_MnI}. Therefore, most of Mn atoms at the interstitial sites %diffused into GaAs 
are presumably located in the tetrahedral interstitial site and surrounded by four As atoms, as in the case of Mn in the substitutional site (Mn$_\textup{\scriptsize{Ga}}$). Because the PES spectra of the transition-metal in the host semiconductor are sensitive to the ligand atoms, the spectra of Mn$_\textup{\scriptsize{I}}$ tetrahedrally coordinated by As atoms will be similar to those of Mn$_\textup{\scriptsize{Ga}}$ in Ga$_{1-x}$Mn$_x$As and Mn in ZB-type MnAs dots, which have same ligand atoms. 
%We note that some of diffused Mn ions are thermally activated and can substitute Ga.

\begin{figure}[tbp]
\begin{center}
\includegraphics[width=8.7cm]{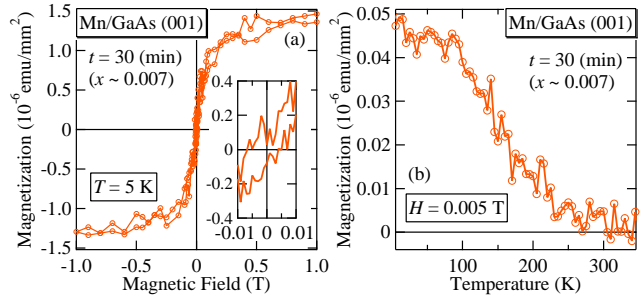}
\caption{(Color online) Magnetization curves of thermally diffused Mn/GaAs (001) interface sputtered for 30 min. Mn concentration is estimated to be $x$ $\sim$0.007 from Fig.\,2(b). The linear component in the high magnetic field region has been subtracted. (a) $M$-$H$ curve at $T$ = 5 K.  The inset shows an enlarged plot from -0.01T to 0.01T. (b) $M$-$T$ curve for $H$ = 0.005 T (zero-field cooled).}
\label{MPMS}
\end{center}
\end{figure}

%\subsection{MAGNETIZATION MEASUREMENTS}
We also performed a magnetization measurement for the sample which had been sputtered for 30 min using the same ion gun as that used for the XPS measurements. Mn concentration is estimated to be $x$ $\sim$0.007 from Fig.\,2(b). The linear component in the high magnetic field region has been subtracted. Fig.\,4(a) is $M$-$H$ curve at $T$ = 5 K.  The inset shows an enlarged plot. A tiny hysteresis was observed, indicating that the sample exhibited ferromagnetic behavior in the dilute Mn phase ($x$ $\sim$0.007). We note that the $T\mathrm{_C}$ estimated from the $M$-$T$ curve shown in Fig.\,4(b), is about 270 K, much higher than that of Ga$_{1-x}$Mn$_x$As ($\sim$110 K, as-grown) but close to that of ZB-type MnAs dots ($\sim$280 K) \cite{Ono_MnAs, Okabayashi_MnAs2}. Furthermore, the $M$-$T$ curve has a long tail (100 K $\sim$270 K) which differs from that of a single component ferromagnet. These features can be explained by a superposition of the $M$-$T$ curves with various $T\mathrm{_C}$ . It is possible that ZB-type MnAs dots with various sizes and/or inhomogeneous distribution, were formed by thermal annealing and exhibited the present ferromagnetic behaviors.

\section*{IV. SUMMARY}
We have performed a depth profile study of thermally diffused Mn/GaAs (001) interfaces using PES combined with Ar$^+$-ion sputtering to investigate the electronic structure of the sample as a function of distance from the surface. We have confirmed that Mn was thermally diffused deep into the GaAs substrate, and completely reacted with GaAs. %, consistent with the large diffusion coefficient reported in previous studies. 
The Mn 2$p$ core-level and Mn 3$d$ valence-band spectra of thermally diffused Mn/GaAs (001) interfaces in the dilute Mn phase were similar to those of Ga$_{1-x}$Mn$_x$As, ZB-type MnAs dots, and/or Mn$_\textup{\scriptsize{I}}$ in tetrahedrally coordinated by As atoms, indicating that the Mn atoms diffused into GaAs do not form any metallic compounds but are tetrahedrally coordinated by ligand As atoms and Mn 3d states are hybridized with the ligand orbitals. Ferromagnetism was observed in the dilute Mn phase.

\section*{ACKNOWLEGDEMENTS}
We thank T.\,Okuda, A.\,Harasawa, and T.\,Kinoshita for their valuable technical support. his work was supported by a Grant-in-Aid for Scientific Research in Priority Area gCreation and Control of Spin Currenth(19048012) from MEXT, Japan. The experiment at Photon Factory was done under the approval of the Photon Factory Program Advisory Committee (Proposal No.\,2004G002).

\end{document}